# Localization of macroscopic sources of magnetic field using optical fibers doped with NV-rich sub-micron diamonds and zero-field resonance

Mariusz Mrózek[1,a], Adam Filipkowski[2,3], Wojciech Gawlik[1], Ryszard Buczyński[2,3], Adam M. Wojciechowski[1], Mariusz Klimczak[2,a]

[1]Institute of Physics, Jagiellonian University in Kraków, Łojasiewicza 11, Kraków 30–348, Poland

[2]Faculty of Physics, University of Warsaw, Pasteura 5, Warsaw 02–093, Poland

[3]Łukasiewicz Research Network - Institute of Microelectronics and Photonics, Al. Lotników 32/46, Warsaw 02–668, Poland

[a]Authors to whom correspondence should be addressed: mariusz.mrozek@uj.edu.pl, mariusz.klimczak@fuw.edu.pl

**Abstract**

We employ an optical fiber doped with randomly oriented fluorescent sub-micron diamonds and the novel zero-field resonance protocol to collect information on the localization and orientation of a magnetic-field source and its distribution. Many previous demonstrations of diamond-based magnetic field sensing achieved ultrahigh sensitivities down to the fT range warranted by manipulating spin states of the diamond nitrogen vacancy (NV) centers with externally applied radio or microwaves. The application of such oscillating fields is problematic in distributed magnetic-field measurements and may be incompatible with specific targets. Instead of relying on these approaches, we leveraged cross-relaxations of particular spin-state populations of the NV center under a magnetic field, thus observing zero-field resonances and making external radio frequency fields redundant. Combined with an optical fiber sensitive to the magnetic field along its entire length, remote sensing was realized that returned information on the spatial field distribution without using any moving mechanical elements in the detection system. Variation of the spatial parameters of the investigated field was achieved simply by controlling the current in a pair of induction coils easily integrable with optical fibers without limiting the fiber-specific functionality of the optical readout taking place at a fixed location at the optical fiber output. Lifting of the requirements related to the mechanical scanning of the fiber, the application of external fields, and the orientation of the NV centers against the measured field mark a very practical step forward in optically driven magnetic field sensing, not easily achievable with earlier implementations.

**Introduction**

Disruption of spin polarization in nitrogen-vacancy defects (NVs) of diamond is a powerful method for sensing and measuring subtle magnetic fields. Combined with alignment of the NV center-symmetry axis with the magnetic-field vector, it allows for vector measurements of the magnetic field. Among the most common techniques is optically detected magnetic resonance (ODMR) [1] which can achieve magnetic field sensitivity significantly better than $1\ nT/\sqrt{Hz}$ [2].

Although impressive, these results come at the price of ODMR-specific limitations, including the incompatibility of some targets with microwaves, such as sensitive biological, magnetic, or superconducting materials. They are also usually restricted to stationary targets of small dimensions, whereas macroscopic stereometric measurements require complex pre-orientation schemes [3]. Diamond's NV center spins can also be manipulated with radio frequency (RF) pulses, which in a recent demonstration unlocked the sensitivity to the magnetic field at the fT level [4]. However, the application of an intense RF field may also be unacceptable for some targets, such as biosamples. A promising solution to the latter limitation has been proposed by exploiting the cross-relaxation of the NV center spin states under scanning with an external magnetic field of a particular orientation [5]. In this approach, scanning the magnetic field along the crystallographic axes of the NV centers can also provide information on the magnetic field vector. Unlike in the ODMR, the ground-state splitting of the NV center spin states is not dependent on temperature in this sensing modality, although cryogenic cooling served to improve the detection of cross-relaxation resonances among differently oriented NV centers under small magnetic field strengths. In a recent study, this near-zero magnetic field resonance technique has been extended over a temperature range of 20 to 250 K, demonstrating the potential of the method for optimization toward sensitivities better than 100 $\frac{pT}{\sqrt{Hz}}$ [6].

Authors in Ref. [7] have demonstrated that the application of nano- and/or microdiamonds (sizes between 30 nm and 3 μm) enables the development of practical diamond-based magnetic field sensors that operate without any oscillating magnetic field. These results also included the evolution of the $T_1$ spin relaxation time versus the particle size, providing evidence that the cross-relaxation between the spin states of NV centers near the zero magnetic field was the sole mechanism behind the observed fluorescence dynamics. Furthermore, it allowed the conclusion that the zero magnetic field fluorescence features did not depend on the diamond orientation, which is an important implication in the combination of the zero-field resonance (ZFR) modality with the volumetric functionalization of optical fibers with nanodiamonds for sensing [8,9].

We extended the ZFR protocol by combining it with optical fibers sensitive to the magnetic field at every point along the propagation length [8,9]. These fibers are distinct in the way that NV-rich sub-micron diamond particles are distributed in the volume of the optical fiber core, in contrast to placing a single diamond crystal at the input facet or in a tapered section, as in previous realizations [3,10]. The diamond orientations in such fibers are random, which precludes a direct vector measurement of the magnetic field. Green pump laser light propagating in the optical fiber core containing diamond particles excites all of their NV center population at once, which is detrimental to the contrast observed in regular ODMR measurements [9]. Because of the spread of diamonds throughout the core, the fiber theoretically allows for sensing magnetic fields distributed along its length. However, physical demonstrations of such measurements have so far been scarce. An example involved an ODMR-based hybrid optofluidic fiber [11]. In that work, the magnetic field measurement was realized with a single droplet with microdiamonds pushed through an optofluidic channel of the fiber. In our previous work employing ODMR [12], we showed that spin information could be transmitted over the length of a diamond-doped optical fiber despite the presence of significant fluorescence background from NV centers that are not subjected to the magnetic or RF field. Unlike our previous work, in this research we set out to demonstrate distributed optical magnetic field sensing without any RF field, with the added functionality of tracking the position and orientation of the macroscopic "source" of the magnetic field, because we control the spin through ZFR.

**Experimental setup and procedure**

The experimental setup (Fig. 1a) consists of a 532-nm laser light source (Sprout-G) with power set to 15 mW for the experiment. The beam is directed to a microscope objective (Motic, UPlan 40×), which couples the light into a fiber optic core. The fluorescence is collected from the other end of the optical fiber by an identical objective and the fluorescence is filtered from the laser light by a filter (Thorlabs, FEL0600) placed in front of the camera (Basler, acA2440). Coils 1 and 2, each with a field/current conversion coefficient of 1.12 mT/A, are connected to a programmable current source (R&S, HMP4040). Control and acquisition were achieved using a computer program written in Python.

In Fig. 1b, we show confocal microscope images of the optical fiber with volumetrically integrated diamonds in the fiber core. The bright red dots mark the locations of the fluorescent diamonds rich in NV$^-$. Observable deformations in the shape of the imaged fluorescent diamonds are not inherent in the particles themselves, but are the result of astigmatism within the imaging system caused by a difference in the refractive index of $\Delta n = 0.09$ between the fiber and the immersion oil surrounding it. The fiber core is fabricated using a dip-coating method to deposit diamond particles on a glass rod, which is then stacked-and-drawn into fibers, a process first demonstrated by the authors in Ref. [8]. In our modified procedure, outlined in detail in ref. [9], we use soft glasses similar to F2 silicate glass (Schott) along with sub-micrometer NV-rich diamonds with a mean particle size of 750 nm (MDNV1um, Adámas Nanotechnologies). The resulting fiber is multimode, with a core diameter of 50 μm and an outer diameter of 125 μm.

The experimental procedure aimed to record spatially resolved data on the position and orientation of an external magnetic field source, a neodymium magnet (1 cm × 2 cm × 0.5 cm), relative to the magnetically sensitive optical fiber. The magnet was moved horizontally stepwise along a coordinate perpendicular to the direction defined by the stretched fiber, on a motorized translation stage (VT-80, PI, USA). Neither the fiber nor the magnet changed their vertical coordinates during the measurement. The measurement sequence was as follows: for each position of the permanent magnet relative to the optical fiber, a camera image of the fiber output facet was taken for the given intensity of the current flowing through coil 1 or coil 2. The current intensity was scanned, which corresponded to variations in the magnetic field intensity generated by the respective coil in the range of –5 mT to 5 mT with steps of 0.1 mT. Separate sets of fiber output facet images (one image per coil, per each current intensity setting) have been recorded for each of the two coils within this magnetic field range. A series of 101 images was used to reconstruct a complete trace of the ZFR signal by varying the current intensity for each coil. Each signal data point for a given step of the magnetic field intensity was retrieved by averaging 16 consecutive images. In post-processing, 16×16 pixels were binned to improve signal quality. Figure 1c shows an example of a ZFR trace reconstructed from images recorded using the described procedure. The trace corresponds to a specific position of the magnet relative to the optical fiber and the current intensity scans in one coil, which generated a magnetic field intensity in the range of –5 mT to 5 mT. Each such ZFR trace was fitted with a Lorentzian function to retrieve the center position (shift), the ZFR depth (contrast), and the resonance width (full width at half-maximum) of the fitted function. The nonzero shift of the ZFR trace reveals a nonzero magnetic field in specific binned pixels and demonstrates the idea of our localization method.

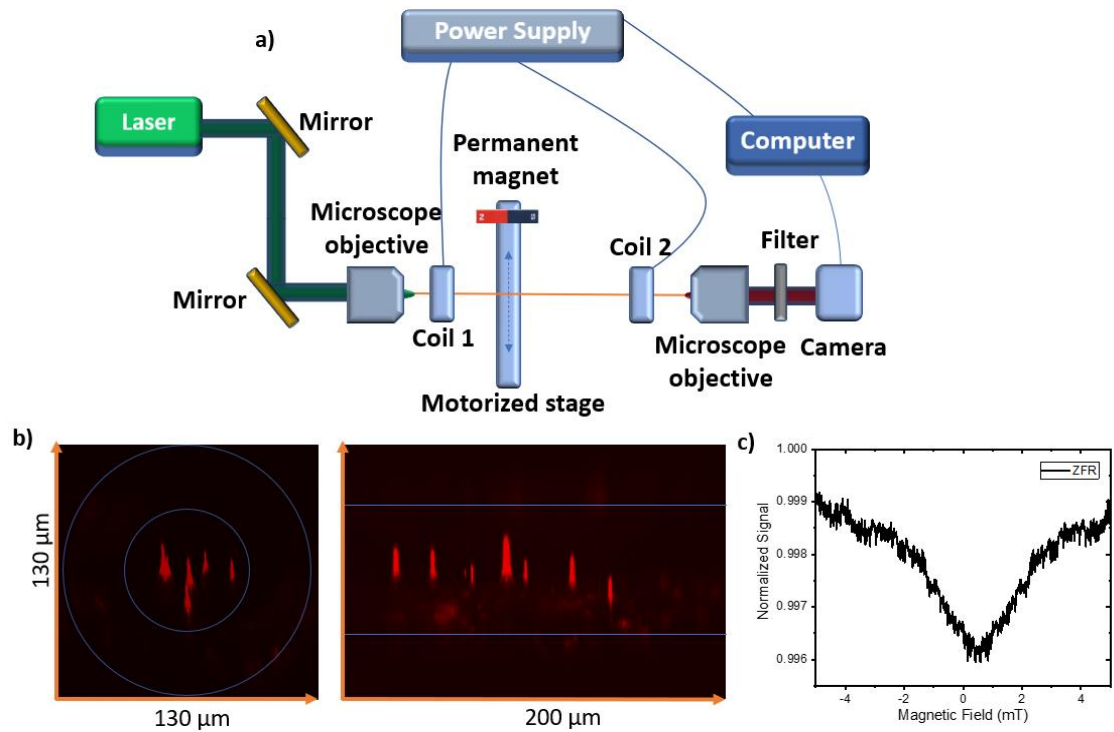

*Fig. 1. a) Experimental setup. b) Confocal microscope images of a section of the developed fiber, diamond fluorescence recorded for the front plane (left), and the side plane (right). c) An example of a zero-field resonance signal for one binned pixel, with a shift of around 0.5 mT.*

**Results and discussion**

Figure 2 shows examples of images of the optical fiber output facets and the corresponding distributions of values of each of the three relevant ZFR parameters when the distance between the permanent magnet and the optical fiber was 15 mm. The fiber core in which the NV-rich sub-micron diamonds are embedded was encircled with a blue trace during image postprocessing in Figs. 2a,b,c.

The signal outside the fiber core area arises from light leakage from the fiber core into the surrounding fiber cladding. This justifies our use of a camera instead of a simple integrating power sensor in these measurements: the camera enables straightforward discrimination of the leakage signal and focuses only on the signal propagating along the fiber core for the determination of the ZFR parameters.

Figures 2d,e,f show these three parameters fitted across a transverse coordinate, that is, the diameter of the fiber core. Note that in each of the three determined ZFR parameters, their distribution within the fiber core is nearly uniform. This allows one to average over the entire fiber core to retrieve the relevant parameter values. This is crucial because the width and contrast of a ZFR trace characterize the magnetic field intensity and sensitivity, while the ZFR shift carries information on both the location and orientation of the magnet against the optical fiber (here it is assumed that both coils are aligned with the direction determined by the length of the straight fiber).

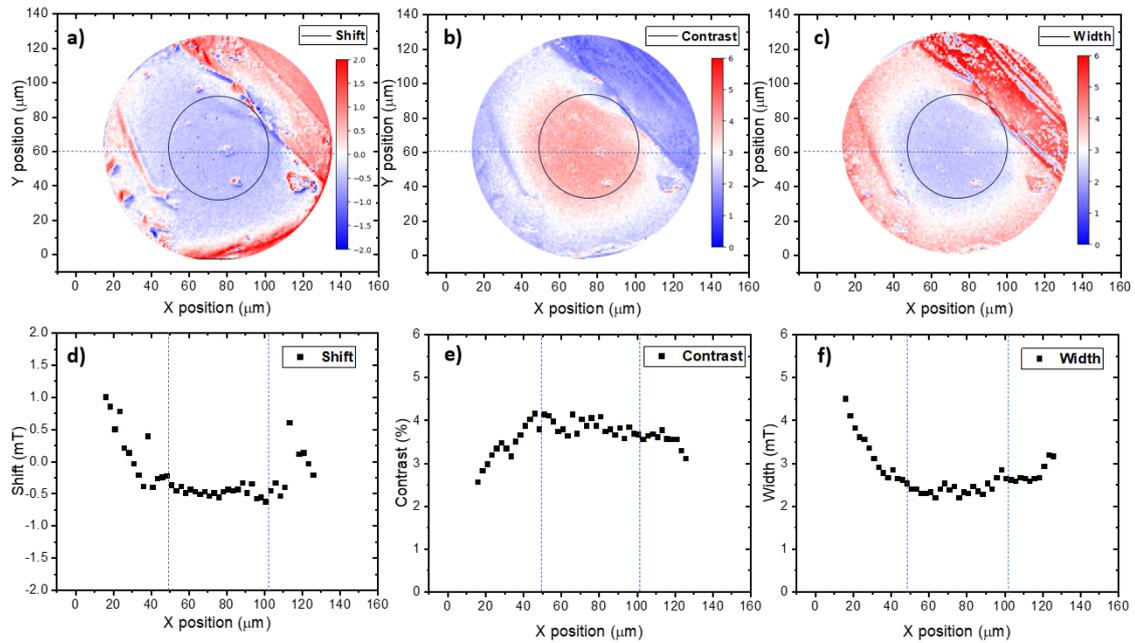

*Fig. 2 Example maps of the ZFR parameter at the output facets of the optical fiber showing a uniform distribution over the core cross section for: a) ZFR shift, b) ZFR contrast, and c) ZFR width. Data were taken by fitting a Lorentzian function to the ZFR traces with current flowing through coil 2. Cross-sectional plots as functions of the position of X for the position Y = 60 um, d) resonance shift, e) contrast, f) width.*

Figures 3a, b, and c illustrate how the position and orientation of the magnet affect the signals acquired along the fiber length by scanning the coil currents flowing through coil 1 or coil 2 for different positions of the magnet over ±80 mm range in the orthogonal direction to the straightened fiber. The perpendicular axis of the magnet movement crosses the fiber at a fixed distance of 70 mm from coil 1. The measurement can be repeated for any value of this distance to mimic the arbitrary distribution of the unknown field to be determined. In addition, the orientation of the magnet poles against the direction of the straight fiber also influences the retrieved ZFR shifts, as shown in the consecutive plots in Fig. 3. These were obtained with the magnet scanned perpendicular to the fiber, but in each of the three cases with its poles at a different angle against the fiber (the angle was fixed during each of the three measurements), as shown in the schematics under each plot. The condition that the fiber remains straight applies to its section located between the two coils. This is motivated by providing a uniform basis of the magnetic field vector space between these coils. The vector basis is defined by the parallelism of the adjacent coils. Indeed, the results here relate to the simplest case involving only two coils, but the method is readily scalable by adding extra coils. Each following pair of adjacent coils would extend the spatial information data set on the location of the magnetic field source and enhance the accuracy of its retrieval. Admittedly, the experiment to retrieve this information would be limited by the rate of electric switching

from coil to coil, the rate of current scanning in an individual coil, and finally by the rate of data acquisition. The method is also scalable in such a way that the distance between two adjacent coils can be selected to match the expected size of the magnetic field source.

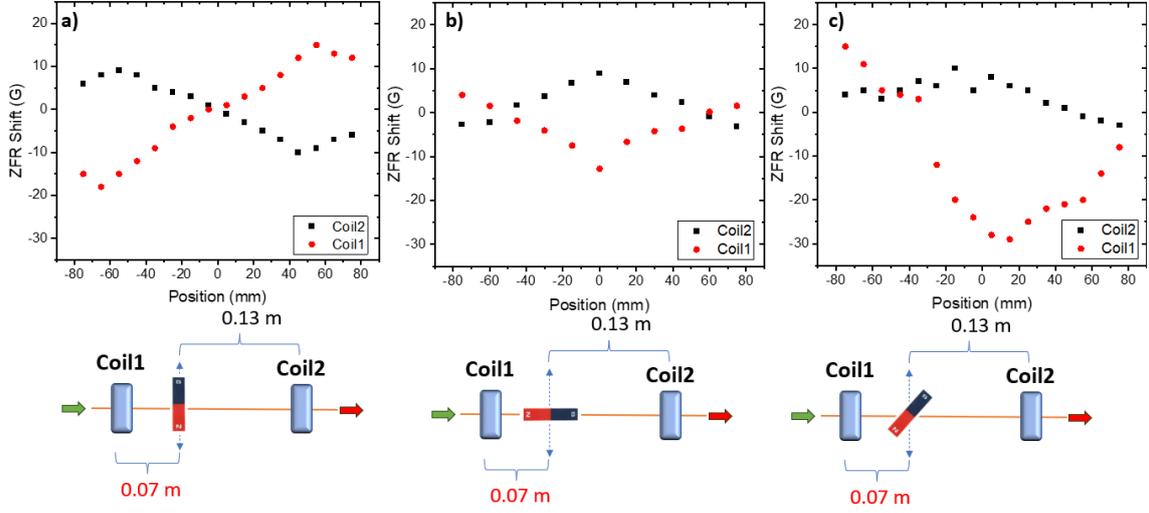

*Fig. 3. ZFR shifts for different positions of the permanent magnet (from -80 to +80 mm perpendicular to the straight fiber at a 70 mm distance from coil 1 and 13 cm from coil 2). Each plot contains two ZFR shift traces for the current flowing through each coil; the red dots represent the signals recovered with the current flowing through coil 1 and the black dots through coil 2. The magnet pole orientation is: a) perpendicular, b) parallel, and c) at 45 deg. relative to the fiber, as illustrated below the plots.*

The magnetic field sensitivity is influenced by various factors, including the ND size and concentration, the orientation of the applied magnetic field, light collection efficiency, and the intensity of the pumping laser. It is typically estimated for a photon-shot-noise-limited scenario [13] using the formula:

$$\eta \approx F \frac{h}{g\mu_B} \frac{\delta f}{C\sqrt{N}} \quad (1)$$

where $\delta f$ denotes the resonance linewidth, $C$ (~$2\times10^{-2}$) is the fractional resonance contrast, $N$ (around $9 \times 10^6$ counts per second for an effective pixel area of $3.5 \times 3.5$ μm$^2$) denotes the number of photons detected, and the prefactor $F$ equals $4/(3\sqrt{3})$ when the resonance lineshape is Lorentzian. In this work, we determined that the per-pixel sensitivity was 4.5 μT/√Hz per μm$^2$ in the case of magnetic fields around 2 mT, which is close to the measurements conducted using the ODMR technique [9].

**Conclusions**

The main achievement reported in this letter was obtaining spatially resolved information on the localization of a magnetic field source in a mechanical scanning-free configuration. An optical fiber doped with NV-rich diamonds has been used similarly to Ref. [12], but here the external MW field has been replaced with probing by an additional magnetic field controlled by external coils. This constituted a complex magnetic field environment where different magnetic objects were operating, each emanating its own field. Despite this ambiguous magnetic field landscape, the fiber sensor enabled dynamic tracing of the location change of the magnetic field source and the orientation of its poles. To precisely determine the location of the magnetic field source, it is necessary to calibrate the system for the given configuration of the coils 1 and 2. In a real-life situation of a non-point magnetic field source, the position reading will be affected by the geometry of the source object, as demonstrated for a random orientation of the poles, as shown in Fig. 3c. Additionally, in a setup limited to just two coils, like in the currently discussed demonstration, the retrieval of the magnet location would require prior knowledge on its strength or the orientation of its poles when it was stationary. This limitation is less relevant for a moving source (magnet) or alternatively – when additional coils are included in the setup along the diamond-doped fiber and a multilateration-type algorithm is applied to retrieve the source location.

In a future experiment, optimization of practical parameters, including sensitivity and tracking rate, could be addressed, e.g., by employing phase-sensitive measurements with a lock-in amplifier and detectors with improved sensitivity. The present results have been achieved without an external RF field. Although external fields could be introduced to improve the sensitivity achievable with the reported sensor, scenarios directly benefiting an external field-free operation, e.g., spintronics or mapping of magnetic fields in electronic components, can already be pointed out [14,15]. The obvious advantage of the ZFR is its non-disrupting character and optical readout, which so far has been reserved for implementations that require the application of a MW field [4]. We experimentally demonstrated that the ZFR effect can be utilized in a medium with a large population of NV-rich sub-micron diamonds spread over a macroscopic dimension. NV centers experience optical excitation and varying magnetic fields in different sections of the medium, resulting in varying fluorescence contribution. Moreover, this simple experiment shows that combining a magnetically sensitive optical fiber with the ZFR protocol enables tracking of the location of a magnetic field source along two coordinates, transverse and longitudinal. This approach can also reveal information about the orientation of the magnetic poles relative to the direction of the optical fiber.

**Funding**

This research was supported by the TEAM NET programme of the Foundation for Polish Science co-financed by the European Union under the European Regional Development Fund, project POIR.04.04.00–00-1644/18 and the study was also carried out using research infrastructure purchased with the funds of the European Union in the framework of the Smart Growth Operational Programme, Measure 4.2; Grant No. POIR.04.02.00-00-D001/20, "ATOMIN 2.0 - ATOMic scale science for the INnovative economy.

**Disclosures**

The authors have no conflicts to disclose.

**Data Availability**

The data that support the findings of this study are available from the corresponding author upon reasonable request.